\documentclass[a4paper,12pt]{article}

\usepackage{amsmath}
\usepackage{amssymb}
\usepackage{color}
\usepackage{mathrsfs}
\usepackage{graphicx}

\DeclareMathOperator{\Li}{Li}

\def\det{\mathrm{det}}
\def\determinant#1{\underset{#1}{\mathrm{det}}}

\def\={\stackrel{\bullet}{=}}

\def\({\left(}
\def\){\right)}
\def\[{\left[}
\def\]{\right]}

\def\cN{{\cal N}}
\def\cO{{\cal O}}

\def\mf {\mathfrak }

\def \be {\begin{equation}}
\def \ee {\end{equation}}
\def \beal#1 {\begin{align}#1\end{align}}
\def \bes#1 {\begin{equation}\begin{split}#1\end{split}\end{equation}}
\def \nn {\notag\\}

\usepackage{geometry}
\numberwithin{equation}{section}
\allowdisplaybreaks

\begin{document}

\newcommand{\hiduke}[1]{\hspace{\fill}{\small [{#1}]}}
\newcommand{\aff}[1]{${}^{#1}$}
\renewcommand{\thefootnote}{\fnsymbol{footnote}}

\begin{titlepage}
\begin{flushright}
{\footnotesize KIAS-P17045}\\
{\footnotesize YITP-17-64}
\end{flushright}
\begin{center}
{\Large\bf
Complete factorization in minimal $\cN=4$ Chern-Simons-matter theory
}\\
\bigskip\bigskip
{\large Tomoki Nosaka\footnote{\tt nosaka@yukawa.kyoto-u.ac.jp}}\aff{1}
{\large and Shuichi Yokoyama\footnote{\tt shuichi.yokoyama@yukawa.kyoto-u.ac.jp}}\aff{2}\\
\bigskip\bigskip
\aff{1}: {\small
\it School of Physics, Korea Institute for Advanced Study\\
85 Hoegiro Dongdaemun-gu, Seoul 02455, Republic of Korea
}\\
\bigskip
\aff{2}: {\small
\it Yukawa Institute for Theoretical Physics, Kyoto University\\
Kitashirakawa-Oiwakecho, Sakyo, Kyoto 606-8502, Japan
}
\end{center}

\begin{abstract}

We investigate an $\cN=4$ $\text{U}(N)_k\times \text{U}(N+M)_{-k}$ Chern-Simons theory coupled to one bifundamental hypermultiplet by employing its partition function, which is given by $2N+M$ dimensional integration via localization. 
Surprisingly, by performing the integration explicitly we find that the partition function completely factorizes into that of
the pure Chern-Simons theory for two gauge groups and an analogous contribution for the bifundamental hypermultiplet.
Using the factorized form of the partition function we argue the level/rank duality, which is also expected from the Hanany-Witten transition in the type IIB brane realization.
We also present the all order 't Hooft expansion of the partition function and comment on the connection to the higher-spin theory.

\end{abstract}

\end{titlepage}

\renewcommand{\thefootnote}{\arabic{footnote}}
\setcounter{footnote}{0}

\tableofcontents
\section{Introduction}
\label{Intro} 

It has been more than a decade since a three dimensional maximally superconformal Chern-Simons-matter theory has been discovered. 
The construction of such a theory was highly important since they are expected to describe {multiple M2-branes} in M-theory \cite{Kitao:1998mf,Bergman:1999na,Schwarz:2004yj}.
The first example was obtained by generalizing the usual Lie algebra to the three bracket to describe ``Chan-Paton'' degrees of freedom associated with the M2-branes \cite{Bagger:2006sk,Bagger:2007jr,Bagger:2007vi,Gustavsson:2007vu} (see also \cite{Basu:2004ed}), which turned out to describe at most two M2-branes in a countable case.
Subsequently superconformal Chern-Simons theories with eight supercharges were constructed as linear quiver gauge theories \cite{Gaiotto:2008sd} associated with type IIB brane configuration \cite{Hanany:1996ie}, without relying on the three bracket.
Then it was insightfully pointed out that by changing the minimal linear quiver gauge theory to the circular one the system achieves supersymmetry enhancement and describes interaction of arbitrary number of membranes \cite{Aharony:2008ug}.\footnote{
For extensions of these quiver theories, see \cite{Fuji:2008yj,Hosomichi:2008jd,Benna:2008zy,Hosomichi:2008jb,Imamura:2008nn}.
}
This opened up the study of superconformal Chern-Simons-matter theories from AdS/CFT correspondence \cite{Maldacena:1997re,Witten:1998qj,Gubser:1998bc}.\footnote{
For some related early studies, see \cite{Bhattacharya:2008bja,Arutyunov:2008if,Nishioka:2008gz,Imamura:2008ji}.
}

One of the non-trivial checks for the AdS/CFT correspondence was given by the exact computation of the partition function with the help of the supersymmetric localization technique \cite{Witten:1988ze,Nekrasov:2002qd,Pestun:2007rz}.
The localization technique reduces the path integral for the partition function to a finite dimensional matrix integration \cite{Kapustin:2009kz,Jafferis:2010un,Hama:2010av}.
These matrix models can be evaluated by the saddle point approximation in the large $N$ limit \cite{Drukker:2010nc,Herzog:2010hf}, and found to be consistent with the classical approximation in the gravity side.
It also provides non-trivial checks for various field theory dualities as some simple integration identities \cite{Kapustin:2010xq,Kapustin:2010mh,Dey:2011pt,Dey:2013nf}.
Moreover, for some special class of ${\cal N}=4$ quiver superconformal Chern-Simons theories the matrix model can be regarded as the partition function of $N$ free fermions in one dimensional space \cite{Marino:2011eh}, which enables us to evaluate the all order $1/N$ corrections explicitly \cite{Calvo:2012du,Hatsuda:2012dt,Putrov:2012zi}.
This lead us to the discovery of the coincidence between the large $N$ instanton effects in the partition function and a non-perturbative completion of the topological string free energy \cite{Hatsuda:2013oxa}, which was not clear a priori from neither the construction of the theory nor the gravity dual.

In this situation we revisit the superconformal Chern-Simons theories characterized by linear quiver diagrams, which are also called Gaiotto-Witten theories \cite{Gaiotto:2008sd}, from its exact partition function determined by the supersymmetric localization.
Particularly in this paper we intensively study the partition function of the minimal $\cN=4$ Chern-Simons theory with U(N)$_k\times$ U$(N+M)_{-k}$ gauge group interacting with a bifundamental hypermultiplet.
Interestingly, we find that the partition function can be decomposed into a product of the pure Chern-Simons partition functions for the two gauge nodes and an analogous factor for the bifundamental representation in $\text{U}(N)\times \text{U}(N+M)$.
We shall refer to this decomposition as {\it complete factorization}.

Our result allows us to argue the level/rank duality of the $\text{U}(N)_k\times \text{U}(N+M)_{-k}$ linear quiver theory, whose evidence was provided in \cite{Yokoyama:2013pxa} in the vector model limit.
We can provide the dictionary of particular observables under the duality by turning on the FI mass deformations.
Once we obtain the partition function in a closed form expression we can also evaluate it in various limits directly.
In particular, written in the form of complete factorization, the all order 't Hooft expansion for the free energy can be obtained easily.
Since the vector model limit corresponds to a boundary of the parameter space of the two 't Hooft couplings, we can discuss the relation of the $\text{U}(N)_k\times \text{U}(N+M)_{-k}$ Gaiotto-Witten theory to the dual higher spin theory, supergravity and their interpolation.
% interpretation.

The rest of this paper is organized as follows.
In the next section we review the basic properties of the $\text{U}(N)_k\times \text{U}(N+M)_{-k}$ linear quiver theory.
In \S\ref{ExactPf}, starting from the localization formula, we write down the partition function of this theory with mass deformation in a closed form expression for general $k$, $N$ and $M$.
We observe two novel structures of the partition function: the complete factorization \eqref{factorize} and the singularity in the massless limit (see \S\ref{commentonbound}).
The complete factorization enables us various subsequent analysis.
In \S\ref{Duality} we argue the level/rank duality from the symmetry of the partition function, which, regarding the mass parameter as fugacity, also provides the dictionary of some observables under the duality.
In \S\ref{tHooft} we evaluate the free energy in the 't Hooft expansion $N,M,k\rightarrow\infty$ with $\lambda_1=N/k$ and $\lambda_2=(N+M)/k$ kept finite.
We also comment on the vector model limit, which is realized as $\lambda_1=0$.
\S\ref{Discuss} is devoted for discussion and comments on further directions.
This paper also contains two appendices in order to help the derivation of the complete factorization.
\S\ref{A_Formulas} is a collection of determinant formulas used in \S\ref{ExactPf}.
In \S\ref{commentonM1} we explain a part of the computation \eqref{calM1result} in detail.

\section{Minimal $\cN=4$ Chern-Simons theory}
\label{N4Theory} 

In this paper we study the minimal $\cN=4$ Chern-Simons theory, that is, $\text{U}(N)_k\times \text{U}(N+M)_{-k}$ Chern-Simons theory interacting with one bifundamental hypermultiplet.  
This theory was first constructed in \cite{Gaiotto:2008sd} as the simplest example in a class of ${\cal N}=4$ linear quiver Chern-Simons matter theory. 
This theory has $\text{SO}(4)$ R-symmetry and possesses the parity invariance accompanied with the exchange of the two gauge fields when the ranks of the gauge groups are the same.
For general levels and ranks, the system has the generalized parity symmetry, under which the levels as well as the ranks are exchanged.
This system is realized by the following type IIB brane configuration. 
\begin{align}
\includegraphics[scale=0.3]{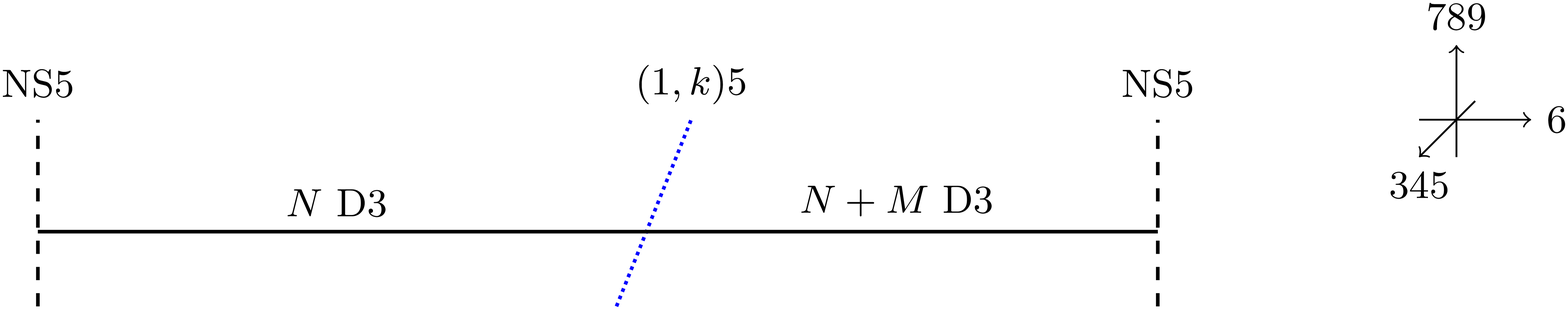}
.
\label{GWN4IIB}
\end{align}
The D3-branes are extending in the $x^6$-direction, ending on the two NS5-branes extending in the $x^7x^8x^9$-directions.
Between the NS5-branes the D3-branes are also intersecting with a $(1,k)$ 5-brane which is extending in the $x^ix^{i+4}$-planes ($i=3,4,5$) with angle $\arctan k$ from each $x^{i+4}$ axis.
In the figure we have omitted $x^0x^1x^2$-directions, where all the branes are extending as well.

It is known that this system admits the mass term keeping $\cN=4$ supersymmetry as well as $\text{SO}(4)$ R-symmetry \cite{Hosomichi:2008jd}. 
In this class of $\cN=4$ theories the mass term is equivalent to FI term. 
In what follows we study the partition function of this system including the $\cN=4$ mass term induced by FI term.
The generalization including the mass term is computationally not difficult but conceptually important.
Indeed, in \S\ref{Duality} we find a non-trivial coincidence of a particular pair of the partition functions by turning on the continuous deformation parameter as well. 
By differentiating the partition function by the deformation parameter in arbitrary times, we obtain the coincidence for many observables.

\section{Exact partition function}
\label{ExactPf}

In this section we study the $S^3$ partition function of the ${\cal N}=4$ $\text{U}(N)_k\times \text{U}(N+M)_{-k}$ Gaiotto-Witten theory with FI deformation. 
With the help of the supersymmetric localization we can reduce the partition function down to the following $2N+M$ dimensional integration \cite{Kapustin:2010xq,Jafferis:2010un,Hama:2010av}
\begin{align}
Z_{\zeta_1,\zeta_2}(k,N,N+M)&=\frac{1}{N!(N+M)!}\int
\Bigl(\frac{d\mu}{2\pi}\Bigr)^N
\Bigl(\frac{d\nu}{2\pi}\Bigr)^{N+M}
\prod_{a=1}^Ne^{\frac{ik}{4\pi}\mu_a^2+i\zeta_1\mu_a}
\prod_{i=1}^{N+M}e^{-\frac{ik}{4\pi}\nu_i^2+i\zeta_2\nu_i}\nonumber \\
&\quad\quad \times \frac{
\displaystyle
\prod_{a<b}^N\Bigl(2\sinh\frac{\mu_a-\mu_b}{2}\Bigr)^2
\prod_{i<j}^{N+M}\Bigl(2\sinh\frac{\nu_i-\nu_j}{2}\Bigr)^2
}
{
\displaystyle
\prod_{a=1}^N\prod_{j=1}^{N+M}2\cosh\frac{\mu_a-\nu_j}{2}
}.
\label{localization}
\end{align}
It is easy to see that this partition function is non-negative when $M=0$, $\zeta_1=-\zeta_2$ and satisfies the generalized parity,
\begin{align}
Z_{\zeta_1,\zeta_2}(k,N,N+M) = Z_{\zeta_2,\zeta_1}(-k,N+M,N).
\label{genpar}
\end{align}

Below we first obtain the closed form expression for $Z_{\zeta_1,\zeta_2}(k,N,N+M)$ for general $k$, $N$, and $M$ by performing the integrations explicitly.
Interestingly, the final expression \eqref{factorize} have completely the same structure as the one-loop determinant factors in the original integrand \eqref{localization}.
That is, the partition function ``factorizes'' into those of the pure Chern-Simons gauge fields and the contributions from the bifundamental matter fields.
We also argue the massless limit $\zeta_1+\zeta_2\rightarrow 0$ of the partition function, where we encounter divergence if the ranks are greater than some critical value determined by $k$.

Note that in the matrix model \eqref{localization} the one-loop determinant factor is not bounded in a generic direction at infinity for general $k$,$N$,$M$, which is in contrast to the case of ABJ(M) matrix model.
Hence the convergence of \eqref{localization} is not trivial.
For the theories without Chern-Simons couplings, it was pointed out \cite{Kapustin:2010xq} that the convergence property of the partition function gives the same classification as that obtained by using monopole operators \cite{Gaiotto:2008sd}.
It would be interesting to examine the convergence of \eqref{localization} rigorously, which will provide a generalization of such classification for the theories including Chern-Simons terms.
In this section, however, we simply assume the convergence and perform the integrations formally to obtain the exact expression for the partition function $Z_{\zeta_1,\zeta_2}(k,N,N+M)$.
Nevertheless the convergence property may be partly reflected in the pole structure of $Z_{\zeta_1,\zeta_2}(k,N,N+M)$.

Even if the partition function is divergent in some parameter regime, we may be able to extract a physical quantity with the help of an appropriate regularization, as is usual in a general quantum field theory.
For example, we can follow the prescription given in \cite{Kapustin:2010xq}.
That is, we can start with the matrix model $Z_{\zeta_1,\zeta_2}(k_1,k_2,N,N+M)$ for the partition function of the $\text{U}(N)_{k_1}\times \text{U}(N+M)_{k_2}$ theory with two independent levels $k_1$ and $k_2$, which is absolutely convergent when $\text{Im}[k_1],\text{Im}[k_2]>0$.
Then the desired partition function $Z_{\zeta_1,\zeta_2}(k,N,N+M)$ can be defined as $\lim_{k_1\to k, k_2\to -k} Z_{\zeta_1,\zeta_2}(k_1,k_2,N,N+M)$, the analytic continuation of $Z_{\zeta_1,\zeta_2}(k_1,k_2,N,N+M)$ as a function of $k_1$ and $k_2$.
Though it is not trivial whether this scheme results in the same expression as \eqref{factorize}, determining the explicit expression of $Z_{\zeta_1,\zeta_2}(k_1,k_2,N,N+M)$ is beyond the scope of this paper hence we shall leave it as a future work.

\subsection{Complete factorization}
\label{CompleteFactor} 

In order to perform the integration \eqref{localization} first we rewrite the one-loop determinant by using the Cauchy-Vandermonde determinant formula \eqref{CauchyVdM} (with appropriate substitutions of $x_a=e^{\mu_a}$, etc.) as follows
\begin{align}
\prod_{a<b}^N2\sinh\frac{\mu_a-\mu_b}{2}&=(-1)^{\frac{N(N-1)}{2}}\det\Bigl[e^{(a-\frac{N+1}{2})\mu_b}\Bigr],
\nonumber \\
\frac{\displaystyle
\prod_{a<b}^N 2\sinh\frac{\mu_a-\mu_b}{2}
\prod_{i<j}^{N+M} 2\sinh\frac{\nu_i-\nu_j}{2}
}
{
\displaystyle
\prod_{a=1}^N
\prod_{j=1}^{N+M}
2\cosh\frac{\mu_a-\nu_j}{2}
}
&=(-1)^{MN+\frac{M(M-1)}{2}}
\determinant{(a\oplus \ell),j}
\begin{pmatrix}
\Biggl[\displaystyle \frac{e^{\frac{M(\mu_a-\nu_j)}{2}}}{2\cosh\frac{\mu_a-\nu_j}{2}}\Biggr]_{N\times (N+M)}\nonumber \\
\Bigl[\displaystyle e^{(\ell-\frac{M+1}{2})\nu_j}\Bigr]_{M\times (N+M)}
\end{pmatrix}\nonumber \\
\prod_{i<j}^{N+M}2\sinh\frac{\nu_i-\nu_j}{2}&=\det\Bigl[e^{(-j+\frac{N+M+1}{2})\nu_i}\Bigr],
\label{CauchyVdMhyperbolic}
\end{align}
where we have used different characters for each type of indices to encode their ranges:
$a,b,\cdots=1,2,\cdots,N$; $\ell,m,\cdots=1,2,\cdots,M$; $i,j,\cdots =1,2,\cdots,N+M$.
By using these the partition function can be rewritten as
\begin{align}
Z_{\zeta_1,\zeta_2}(k,N,N+M)&=\frac{(-1)^{\frac{(N+M)(N+M-1)}{2}}}{N!(N+M)!}\int
\Bigl(\frac{d\mu}{2\pi}\Bigr)^N
\Bigl(\frac{d\nu}{2\pi}\Bigr)^{N+M}
\prod_{a=1}^Ne^{\frac{ik}{4\pi}\mu_a^2+i\zeta_1\mu_a}
\prod_{i=1}^{N+M}e^{-\frac{ik}{4\pi}\nu_i^2+i\zeta_2\nu_i}\nonumber \\
&\quad \times \determinant{a,b}\biggl[e^{(a-\frac{N+1}{2})\mu_b}\biggr]
\determinant{(a\oplus\ell),j}
\begin{pmatrix}
\Bigl[\frac{e^{\frac{M(\mu_a-\nu_j)}{2}}}{2\cosh\frac{\mu_a-\nu_j}{2}}\Bigr]_{N\times (N+M)}\\
\Bigl[e^{(\ell-\frac{M+1}{2})\nu_j}\Bigr]_{M\times (N+M)}
\end{pmatrix}
\determinant{i,j}\biggl[e^{(-j+\frac{N+M+1}{2})\nu_i}\biggr].
\end{align}
Then we can combine the determinants by using the formula \eqref{detdet}, which allows us to move the integration into each matrix element inside a whole determinant.
We finally obtain
\begin{align}
Z_{\zeta_1,\zeta_2}(k,N,N+M)=(-1)^{\frac{(N+M)(N+M-1)}{2}}
\determinant{(a\oplus \ell),j}
\begin{pmatrix}
\Bigl[{\mathscr M}_{1,a,j}\Bigr]_{N\times (N+M)}\\
\Bigl[{\mathscr M}_{2,\ell,j}\Bigr]_{M\times (N+M)}
\end{pmatrix}.
\label{Zindet}
\end{align}
Here the matrix elements of each block are given as
\begin{align}
{\mathscr M}_{1,a,j}&=\int\frac{d\mu}{2\pi}\frac{d\nu}{2\pi}e^{(a-\frac{N-M+1}{2})\mu}e^{\frac{ik}{4\pi}\mu^2+i\zeta_1\mu}\frac{1}{2\cosh\frac{\mu-\nu}{2}}e^{-\frac{ik}{4\pi}\nu^2+i\zeta_2\nu}e^{(-j+\frac{N+1}{2})\nu},\label{matrixelements} \\
{\mathscr M}_{2,\ell,j}&=\int\frac{d\nu}{2\pi}e^{-\frac{ik}{4\pi}\nu^2+i\zeta_2\nu+(\ell-j+\frac{N}{2})\nu},
\notag\end{align}
which are directly computed as follows. 
The second block is easily computed by the gaussian integration as 
\begin{align}
{\mathscr M}_{2,\ell,j}=\frac{e^{-\frac{\pi i}{k}(i\zeta_2- j+ \ell+{N\over2})^2}}{\sqrt{i k}}. 
\label{calM2result}
\end{align}

To compute the first block, we change the integration variables to $\mu_\pm=i(\mu\pm\nu)$
\begin{align}
{\mathscr M}_{1,a,j}&=\int\frac{d\mu_+d\mu_-}{8\pi^2}e^{-\frac{ik}{4\pi}\mu_+\mu_- }\frac{1}{2\cosh\frac{i\mu_-}{2}}e^{(a-\frac{N-M+1}{2}+i\zeta_1){\mu_+ + \mu_- \over 2i} + (-j+\frac{N+1}{2}+i\zeta_2){\mu_+ - \mu_- \over 2i}}.
\end{align}
Thanks to the fact that the Chern-Simons levels add up to zero, the $\mu_+$ integration just gives the delta function and we can easily perform the $\mu_-$ integration as  
\begin{align}
{\mathscr M}_{1,a,j}=&\int\frac{d\mu_-}{4\pi} \delta\Bigl(\frac{k}{4\pi}\mu_- +\frac{(a-j+\frac{M}{2}+i(\zeta_1+\zeta_2) )}{2} \Bigr) \frac{1}{2\cosh\frac{i\mu_-}{2}}e^{(a-\frac{N-M+1}{2}+i\zeta_1+j-\frac{N+1}{2}-i\zeta_2){\mu_- \over 2i}} \nn
=&\frac{e^{\frac{\pi i}{k}(a-\frac{N-M+1}{2}+i\zeta_1)^2}}{k}
\frac{1}{2\cosh\frac{\pi i}{k}(a-j+\frac{M}{2}+i(\zeta_1+\zeta_2))}
e^{-\frac{\pi i}{k}(j-\frac{N+1}{2}-i\zeta_2)^2}.
\label{calM1result}
\end{align}
Substituting these two expressions to \eqref{Zindet} we obtain
\begin{align}
Z_{\zeta_1,\zeta_2}(k,N,N+M)&=(-1)^{\frac{(N+M)(N+M-1)}{2}}\prod_{a=1}^N\frac{e^{\frac{\pi i}{k}(a-\frac{N-M+1}{2}+i\zeta_1)^2}}{k}\prod_{\ell=1}^M\frac{e^{-\frac{\pi i}{k}(\ell-\frac{1}{2})^2}}{\sqrt{ik}}\prod_{j=1}^{N+M}e^{-\frac{\pi i}{k}(j-\frac{N+1}{2}-i\zeta_2)^2}\nonumber \\
&\quad \times \determinant{(a\oplus \ell),j}
\begin{pmatrix}
\Bigl[\frac{1}{2\cosh\frac{\pi i}{k}(a-j+\frac{M}{2}+i(\zeta_1+\zeta_2))}\Bigr]_{N\times (N+M)}\\
\Bigl[e^{\frac{2\pi i}{k}(\ell-\frac{1}{2})(j-\frac{N+1}{2}-i\zeta_2)}\Bigr]_{M\times (N+M)}
\end{pmatrix}.
\label{Zindetfinal}
\end{align}
Here we have factored out some phases in \eqref{calM1result} and \eqref{calM2result}, which depend only on one of the row/column indices and hence can be interpreted as the multiplication of a diagonal matrix.

Notice that the determinant in the last expression \eqref{Zindetfinal} is of the form of the right-hand side of \eqref{CauchyVdMhyperbolic} with the following identification of the parameters\footnote{
One may interpret this result as the partition function being computed with only a single localization locus \eqref{frozenmoduli}.
It would be interesting if one could find another formulation like Higgs branch localization \cite{Fujitsuka:2013fga,Benini:2013yva} to justify this viewpoint.
}
\begin{align}
\mu_a\rightarrow \frac{2\pi i}{k}\Bigl(a-\frac{N-M+1}{2}+i\zeta_1\Bigr),\quad
\nu_j \rightarrow \frac{2\pi i}{k}\Bigl(j-\frac{N+1}{2}-i\zeta_2\Bigr),
% \nu_i\rightarrow \frac{2\pi i}{k}\Bigl(j-\frac{N+1}{2}-i\zeta_2\Bigr),
\label{frozenmoduli}
\end{align}
which allows us to decompose the determinant into the product of hyperbolic functions again.
Collecting the overall factors together, we finally find a surprisingly simple expression for the partition function
\begin{align}
Z_{\zeta_1,\zeta_2}(k,N,N+M)=P Z^{\text{(CS)}}_{\zeta_1}(k,N)Z^{\text{(CS)}}_{\zeta_2}(-k,N+M)Z^{\text{(mat)}}_{\zeta_1+\zeta_2}(k,N,N+M).
\label{factorize}
\end{align}
Here $P$ is a phase factor given by
\begin{align} 
P=e^{-\frac{\pi iM(2 N^2-N (N+M)+2 (N+M)^2-2)}{12k}}i^{-\frac{(N+M)^2+N^2}{2}},
\end{align}
and $Z^{\text{(CS)}}_{\zeta}(k,N)$ is the partition function of pure Chern-Simons theory with FI deformation
\begin{align}
Z^{\text{(CS)}}_{\zeta}(k,N)=\frac{e^{-\frac{\pi iN\zeta^2}{k}}}{k^\frac{N}{2}}\prod_{a>b}^N2\sin\frac{\pi(a-b)}{k},
\label{pureCS}
\end{align}
while $Z^{\text{(mat)}}_{\zeta}(k,N,N+M)$ is given by
\begin{align}
Z^{\text{(mat)}}_{\zeta}(k,N,N+M)=\frac{1}{\displaystyle \prod_{a=1}^{N}\prod_{j=1}^{N+M}2\cos\frac{\pi}{k}\Bigl(a-j+\frac{M}{2}+i\zeta\Bigr)}.
\label{Zmat}
\end{align}
That is, the partition function factorizes into the contributions from the two vector multiplets and the one from the bifundamental hypermultiplet completely.
This is non-trivial since in the original theory the bifundamental hypermultiplet couples to both vector multiplets non-trivially.
Note that, although we have introduced two FI parameters $\zeta_1,\zeta_2$, the ``hypermultiplet contribution'' $Z^{\text{(mat)}}_\zeta(k,N,N+M)$ depends only on a single FI parameter $\zeta$ with $\zeta=\zeta_1+\zeta_2$.
Indeed this dependence corresponds to the mass of the hypermultiplet $(\zeta_1+\zeta_2)/k$, which also motivate us to call the ratio $Z_{\zeta_1,\zeta_2}(k,N,N+M)/(PZ_{\zeta_1}^{\text{(CS)}}(k,N)Z_{\zeta_2}^{\text{(CS)}}(-k,N+M))$ as the contributions from the hypermultiplet $Z^{\text{(mat)}}_{\zeta_1+\zeta_2}(k,N,N+M)$.
It is not difficult to check that this expression of partition function is non-negative when $M=0$, $\zeta_1=-\zeta_2$ and satisfies the generalized parity \eqref{genpar}.

For later convenience we present a different expression for the matter partition function: 
\begin{align}
&Z^{\text{(mat)}}_{\zeta}(k,N,N+M)\nonumber \\
&=\frac{1}{\displaystyle \prod_{\alpha=1}^N\Bigl[2\cos\frac{\pi}{k}\Bigl(\alpha+\frac{M}{2}+i\zeta\Bigr)\cdot 2\cos\frac{\pi}{k}\Bigl(\alpha+\frac{M}{2}-i\zeta\Bigr)\Bigr]^{N-\alpha}
\prod_{\alpha=0}^M\Bigl[2\cos\frac{\pi}{k}\Bigl(\alpha-\frac{M}{2}-i\zeta\Bigr)\Bigr]^N}.
\label{rearrangedproduct}
\end{align}

In the above computation, we have naively used the delta function \eqref{calM1result} to perform the integrations in ${\mathscr M}_{1,a,j}$ \eqref{matrixelements}.
Since the kernels of the delta function are not located on the real axis, however, this would require a more careful argument on the deformation of the integration contours, which we explain in \S\ref{commentonM1} in detail.

Unfortunately the explanation in \S\ref{commentonM1} works only in the regime $N+M/2-1<|k|/2$.\footnote{
A similar problem was already pointed out for the circular quiver superconformal Chern-Simons theories with rank deformations \cite{Moriyama:2016xin,Moriyama:2016kqi,Kiyoshige:2016lno}, where we again encounter an obstacle below the s-rule bounds.
}
In the following sections, however, we shall tentatively assume that our result \eqref{Zmat} is valid beyond the region $N+M/2-1<|k|/2$ by the analytic continuation in $k$, and argue the level/rank duality relation in \S\ref{DualityFromPF} from the partition function.
Performing the integration beyond this region is left as a future work.

\subsection{Poles in massless limit}
\label{commentonbound}

Lastly, let us consider the massless limit $\zeta_1+\zeta_2\rightarrow 0$ of the partition function.
In this limit, the partition function diverges for sufficiently large values of ranks $N,M$ due to the cosine factors in the denominator of 
$Z^{\text{(mat)}}_{\zeta}(k,N_1,N_2)$.

Let us clarify the concrete upper bound for the ranks where the partition function diverges at the saturation of the bound.
The divergence occurs when a pair of indices $(a,j)$ satisfies the following condition
\begin{align}
a-j+\frac{k+N_2-N_1}{2}=mk
\label{polecondition}
\end{align}
for some $m\in\mathbb{Z}$.
From this expression we immediately find that there are no divergence for odd $k+N_2-N_1$.
In the case of even $k+N_2-N_1$, to avoid the divergence the range of the left-hand side of \eqref{polecondition} has to be shorter than $|k|$.
As the left-hand side of \eqref{polecondition} is maximized at $(a,j)=(N_1,1)$ and minimized at $(a,j)=(1,N_2)$, we can easily identify this condition as the following inequality
\begin{align}
N_1+N_2<|k|+2.
\end{align}
In summary, the partition function of ${\cal N}=4$ $\text{U}(N)_k\times \text{U}(N+M)_{-k}$ linear quiver superconformal Chern-Simons theory is finite if $(k,N,M)$ satisfies the following constraint
\begin{align}
k+M\in 2\mathbb{N}-1\quad\text{or}\quad N+\frac{M}{2}-1<\frac{|k|}{2}.
\label{Zmatbound}
\end{align}

Note that the upper bound $N+M/2-1=|k|/2$ is below the zero of the pure Chern-Simons partition functions, where the brane configuration \eqref{GWN4IIB} violates the s-rule \cite{Hanany:1996ie} and hence breaks supersymmetry.
Therefore it seems that the partition function can actually be singular inside the regime where the matrix model \eqref{localization} obtained by supersymmetric localization physically makes sense.\footnote{
Precisely speaking, our computation without the regularization discussed in section \ref{ExactPf} collapses before we encounter a pole due to the bound for justification \eqref{noresiduebound}.
Nevertheless our computation indicates that the $\text{U}(N)_k\times \text{U}(N+M)_{-k}$ Gaiotto-Witten theory should be classified by the inequality $N+M/2-1<|k|/2$.
}
We hope to clarify the interpretation for this divergence in future work \cite{NYfuture_bound}.

\section{Level/rank duality}
\label{Duality} 

In this section we argue the level/rank duality of the ${\cal N}=4$ $\text{U}(N)_k\times \text{U}(N+M)_{-k}$ superconformal Chern-Simons theory. 
We provide a strong evidence for this from the partition function exactly computed as \eqref{factorize} by verifying its invariance under the level/rank duality transformation, and from the type IIB brane configuration by moving 5-branes taking into account the Hanany-Witten effect \cite{Hanany:1996ie}.

\subsection{From partition function }
\label{DualityFromPF}  

Let us show the level/rank duality from the partition function. 
Since our partition function is completely factorized \eqref{factorize}, it is sufficient to show the duality for each factor.
We shall ignore the phase factor $P$ independent of the FI parameters since it is irrelevant for the observables generated by $\partial_{\zeta_1}^m\partial_{\zeta_2}^n \log Z_{\zeta_1,\zeta_2}(k,N,N+M)$.

It is famous that the pure Chern-Simons partition function enjoys the level/rank duality
\begin{align}
Z^{\text{(CS)}}_\zeta(k,N)=Qe^{-\pi i\zeta^2}Z^{\text{(CS)}}_\zeta(-k,\zeta,k-N),
\end{align}
with a phase factor $Q=e^{\frac{\pi i(k-N)^2}{2}}$ independent of $\zeta$.
We can show that the matter contribution $Z^{\text{(mat)}}_\zeta(k,N_1,N_2)$ also satisfy a similar self-dual relation 
\begin{align}
Z^{\text{(mat)}}_\zeta(k,N_1,N_2)=R\Bigl(2\sin\pi\Bigl(\frac{k+N_2-N_1}{2}+i\zeta\Bigr)\Bigr)^{k-N_1-N_2}Z^{\text{(mat)}}_\zeta(k,k-N_2,k-N_1)
\end{align}
with a $\zeta$-independent phase factor $R=(-1)^{\frac{(k-N_1-N_2)(k+N_2-N_1-1)}{2}}$.
This relation follows from the multiplication formula
\begin{align}
\prod_{\ell=1}^{k}2\cos\frac{\pi(z-\ell)}{k}&=(-1)^k2\sin\pi\Bigl(z+\frac{k}{2}\Bigr).
\label{sineadditiveformula}
\end{align}
Indeed, by filling one product by the complement $\prod_{j=1}^N=\prod_{j=1}^k/\prod_{j=N_2+1}^k$ we can apply this formula to the product with full range $\prod_{j=1}^k$, which generates a factor with a trivial dependence on the other index $a$.
Therefore, up to overall phases we can replace $N_2\rightarrow k-N_2$ by moving the product from the denominator to the numerator.
Since we have two products, after the operation twice the product of cosine factors finally come back to the denominator together with the replacement $N_1\rightarrow k-N_1$, $N_2\rightarrow k-N_2$, as
\begin{align}
&\frac{1}{\prod_{a=1}^{N_1}\prod_{j=1}^{N_2}2\cos\frac{\pi}{k}(a-j+\frac{N_2-N_1}{2}+i\zeta)}\nonumber \\
&=\frac{1}{\prod_{a=1}^{N_1}(-1)^k2\sin\pi(a+\frac{k+N_2-N_1}{2}+i\zeta)}
\prod_{a=1}^{N_1}\prod_{i=1}^{k-N_2}2\cos\frac{\pi}{k}\Bigl(a+i-k-1+\frac{N_2-N_1}{2}+i\zeta\Bigr)\nonumber \\
&=\frac{1}{\prod_{a=1}^{N_1}(-1)^k2\sin\pi(a+\frac{k+N_2-N_1}{2}+i\zeta)}\frac{\prod_{i=1}^{k-N_2}(-1)^k2\sin\pi(-i+1+\frac{3k-(N_2-N_1)}{2}-i\zeta)}{\prod_{i=1}^{k-N_2}\prod_{b=1}^{k-N_1}2\cos\frac{\pi}{k}\Bigl(-b+i+\frac{N_2-N_1}{2}+i\zeta\Bigr)}\nonumber \\
&=(-1)^{\frac{(k-N_1-N_2)(k+N_2-N_1-1)}{2}}\Bigl(2\sin\pi\Bigl(\frac{k+N_2-N_1}{2}+i\zeta\Bigr)\Bigr)^{k-N_1-N_2}\nonumber \\
&\quad \times \frac{1}{\prod_{i=1}^{k-N_2}\prod_{b=1}^{k-N_1}2\cos\frac{\pi}{k}(i-b+\frac{N_2-N_1}{2}+i\zeta)},
\end{align}
where we have renamed the indices as $i=k-j+1$ in the second line and $b=k-a+1$ in the third line respectively.

Substituting these results into the factorized expression for the partition function \eqref{factorize}, we obtain the following self-dual relation
\begin{align}
Z_{\zeta_1,\zeta_2}(k,N,N+M)&=e^{\pi i(-\zeta_1^2+\zeta_2^2)}
\Bigl(2\sin\pi\Bigl(\frac{k+M}{2}+i(\zeta_1+\zeta_2)\Bigl)\Bigr)^{k-2N-M}\nonumber \\
&\quad Z_{\zeta_2,\zeta_1}(k,k-N-M,k-N),
\label{GWlevelrank}
\end{align}
up to a $\zeta$-independent phase.
We shall call this the level/rank duality for the ${\cal N}=4$ $\text{U}(N)_k\times \text{U}(N+M)_{-k}$ theory.

Now let us consider the massless limit and argue the relation among the observables, regarding both sides of \eqref{GWlevelrank} as the generating function.
First of all we notice that the prefactor in the massless limit $2\sin\pi((k+M)/2)^{k-2N-M}$ is zero or infinity for even $k+M$.
This is indeed consistent with the observation in \S\ref{commentonbound}, as either $Z_{\zeta_1,\zeta_2}(k,N,N+M)$ or $Z_{\zeta_2,\zeta_1}(k,k-N-M,k-N)$ is singular.
Here we shall focus on the case with odd $k+M$.
Since $\sin\pi((k+M)/2+i(\zeta_1+\zeta_2))\propto \cosh\pi(\zeta_1+\zeta_2)$ up to a phase independent of $\zeta_1,\zeta_2$ in this case, we obtain $(m+n\ge 1)$
\begin{align}
&\frac{\partial^{m+n}}{\partial \zeta_1^m\partial \zeta_2^n}\log Z_{\zeta_1,\zeta_2}(k,N,N+M)\biggr|_{\zeta_1=\zeta_2=0}\nonumber \\
&=2\pi i(-\delta_{m,2}\delta_{n,0}+\delta_{m,0}\delta_{n,2})
+(k-2N-M)\partial^{m+n}_\zeta\Bigl[\log 2\cosh \pi\zeta\Bigr]\Bigr|_{\zeta=0}\nonumber \\
&\quad +\frac{\partial^{m+n}}{\partial \zeta_1^n\partial \zeta_2^m}
\log Z_{\zeta_1,\zeta_2}(k,k-N-M,k-N)
\biggr|_{\zeta_1=\zeta_2=0}.
\end{align}

\subsection{From brane realization }
\label{DualityFromBrane}

The level/rank duality of this system can be understood from the type IIB brane realization as shown in other $\cN=3$ or ${\cal N}=4$ superconformal Chern-Simons matter theories \cite{Giveon:2008zn,Kapustin:2010mh}. 

For simplicity, let us start with the massless case, whose brane configuration is given as \eqref{GWN4IIB}.
To consider the Hanany-Witten transition involving $(1,k)$5-brane, we shall follow \cite{Giveon:2008zn}.\footnote{
The Hanany-Witten transition formula involving general $(q,p)$5-branes was also given in \cite{Kitao:1998mf}.}
First we decompose the $(1,k)$ 5-brane into the NS5-brane extending in the $x^7x^8x^9$-directions and $k$ D5-branes extending in the $x^3x^4x^5$-directions, where the D5-branes are slightly on the left of the NS5-brane as
\begin{align}
\includegraphics[scale=0.3]{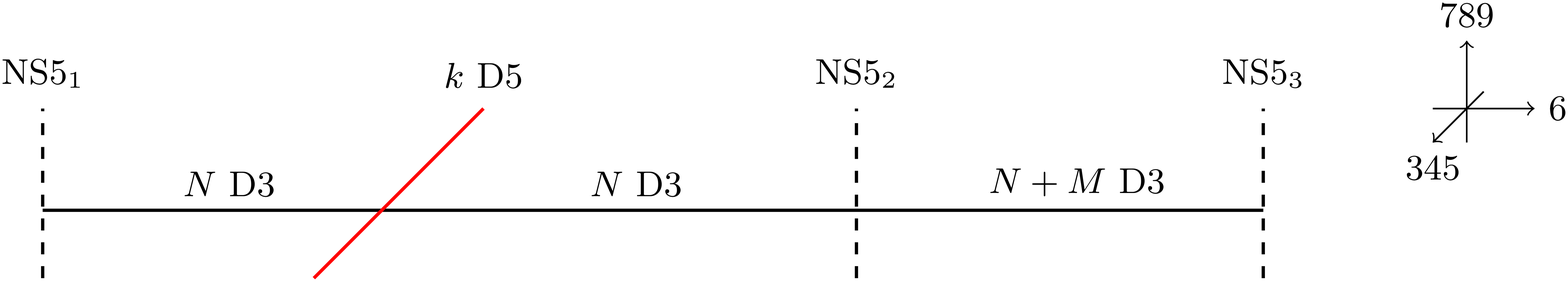}
,
\end{align}
and move the rightmost NS5-brane to the leftmost.
When the 5-branes are exchanged, the number of D3-branes between them is changed so that the {\it linking number} associated with each 5-brane is unchanged \cite{Hanany:1996ie}. Here the linking number for an NS5-brane is defined by 
\be 
\mf L_{\rm NS5} = {R_{\rm D5}-L_{\rm D5} \over 2} +{l_{\rm D3}-r_{\rm D3}} 
\ee
where $R_{\rm D5}[L_{\rm D5}]$ is the number of all D5-branes to the right[left] of the NS5-brane and $r_{\rm D3}[l_{\rm D3}]$ is the number of D3-branes emanating from the NS5-brane to the right[left], and similarly the linking number for a D5-brane is defined by 
\be 
\mf L_{\rm D5} = {R_{\rm NS5}-L_{\rm NS5} \over 2} +{l_{\rm D3}-r_{\rm D3}}, 
\ee
where $R_{\rm NS5}[L_{\rm NS5}]$ is the number of all NS5-branes to the right[left] of the D5-brane and $r_{\rm D3}[l_{\rm D3}]$ is the number of D3-branes emanating from the D5-brane to the right[left]. 
After the rightmost NS5-brane moves to the leftmost, the brane configuration becomes 
\begin{align}
\includegraphics[scale=0.3]{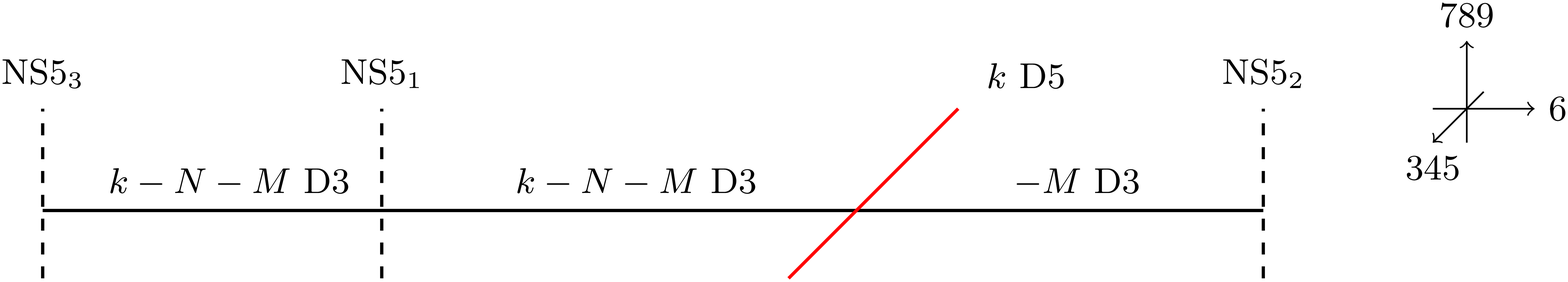}
.
\end{align}
Subsequently we move the NS5-brane at the 1st slot to the rightmost, 
which results in the following brane configuration
\begin{align}
\includegraphics[scale=0.3]{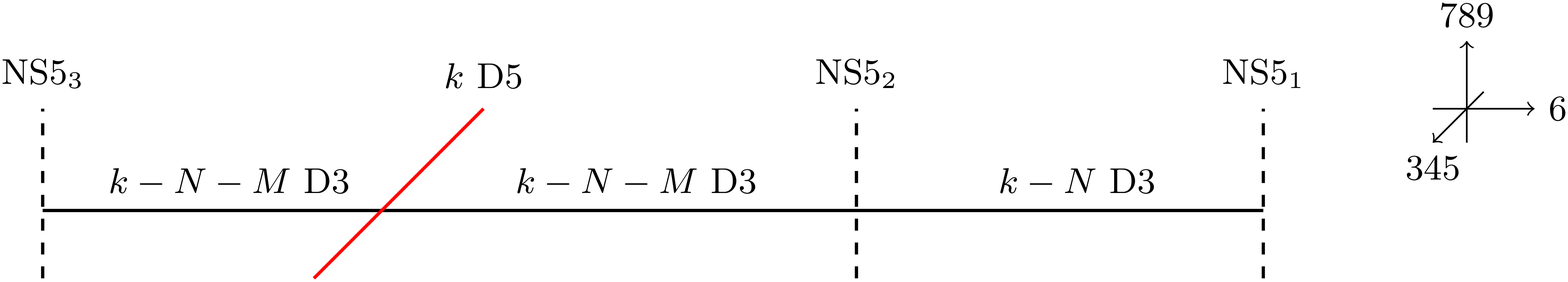}
.
\end{align}
By molding $k$ D5-branes to the middle NS5-brane and regard them again as an $(1,k)$5-brane, we finally obtain 
\begin{align}
\includegraphics[scale=0.3]{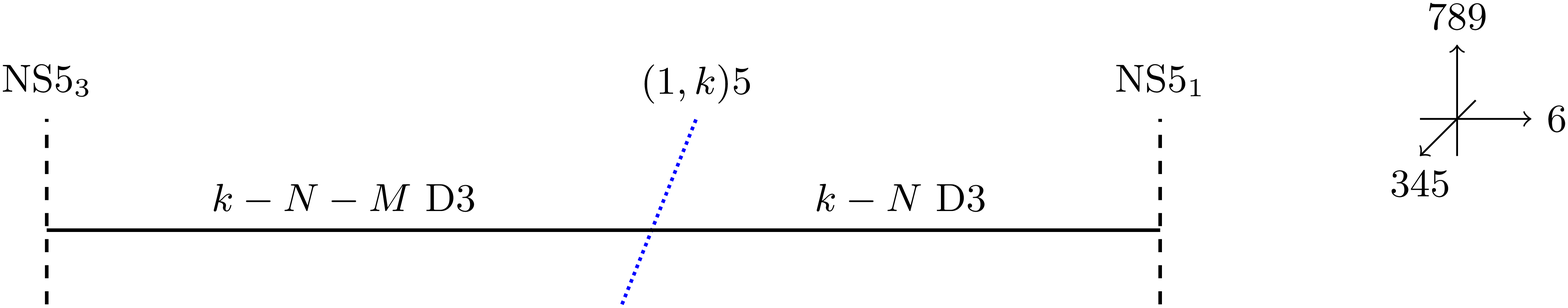}
.
\end{align}
The theory realized on the D3-branes of this brane configuration is $\cN=4$ $\text{U}(k-N-M)_k \times \text{U}(k-N)_{-k}$ Chern-Simons theory, which is the desired result. 

Next let us include the FI terms for this duality argument. 
Following \cite{deBoer:1997ka} we realize the FI term for a gauge group realized on a stack of D3-branes ending with two 5-branes by the difference of the positions of the 5-branes in a transverse direction.
Then $\zeta_1$, $\zeta_2$ are realized by the difference of the positions of the NS5$_1$-brane and NS5$_2$-brane, NS5$_2$-brane and NS5$_3$-brane, respectively.  
After the above interchange of 5-branes, the two FI terms are clearly exchanged with extra minus, which is unimportant in the partition function.  
This completes our argument of level/rank duality from the brane configuration,
up to the overall factor $e^{\pi i(-\zeta_1^2+\zeta_2^2)} (2\sin\pi ((k+M)/2+i(\zeta_1+\zeta_2)))^{k-2N-M}$ in \eqref{GWlevelrank}. 

\section{Free energy in the 't Hooft expansion}
\label{tHooft}
In this section we consider the $1/N$ expansion of the free energy of the ${\cal N}=4$ $\text{U}(N)_k\times \text{U}(N+M)_{-k}$ theory.
To obtain a non-trivial result, we have to take the 't Hooft limit $N,M,k\rightarrow\infty$ with $\lambda_1=N/k$ and $\lambda_2=(N+M)/k$ kept finite, where $\lambda_1$ and $\lambda_2$ are bounded as $0\le \lambda_i\le 1$ due to the s-rule.
As the 't Hooft expansion for the pure Chern-Simons free energy was already studied in full detail (see e.g. \cite{Marinotext,Yokoyama:2016ktm}), let us focus on $Z^\text{(mat)}_{\zeta_1+\zeta_2}(k,N,N+M)$ \eqref{Zmat}, which is the remaining part in the factorization structure \eqref{factorize} and is also of our interest in the vector model limit $M,k\rightarrow\infty$ with $M/k$ and $N$ kept finite.

For simplicity below we consider the massless case $\zeta_1=\zeta_2=0$ and neglect the phases, $F^{\text{(mat)}}(k,N,N+M)=-\log{Z}^{\text{(mat)}}_0(k,N,N+M)$, where the summations coming from the products can be rearranged as (see \eqref{rearrangedproduct})
\begin{align}
F^{\text{(mat)}}(k,N,N+M)=
2\sum_{\alpha=1}^{N}(N-\alpha)\log 2\cos\frac{\pi}{k}\Bigl(\alpha+\frac{M}{2}\Bigr)+N\sum_{\alpha=0}^M\log 2\cos\frac{\pi}{k}\Bigl(\alpha-\frac{M}{2}\Bigr).
\label{rearranged}
\end{align}
Then the large $k$ expansion of $F^{\text{(mat)}}(k,N,N+M)$ can be evaluated in the 't Hooft limit by using the standard technique employed for the pure Chern-Simons partition function \cite{Marinotext,Yokoyama:2016ktm}, which will take the following form of genus expansion
\begin{align}
F^{\text{(mat)}}(k,N,N+M)=\sum_{g\geq 0} F^{\text{(mat)}}_g(\lambda_1,\lambda_2)k^{2-2g},
\label{tHooftExpansion}
\end{align}
with each coefficients $F_g^{\text{(mat)}}(\lambda_1,\lambda_2)$ given in \eqref{Fmatplanar} and \eqref{Fmatatgenusg} below.

Below we first consider the planar limit, $F^{\text{(mat)}}_0(\lambda_1,\lambda_2)$.
Subsequently we determine the all order correction in $1/k$ in an explicit form, where we also assume $M\in 2\mathbb{N}$ for a technical reason.
Finally we further comment on the vector model limit, which corresponds to the limit $\lambda_1\rightarrow 0$ of the planar free energy in our notation.

\subsection{Planar approximation}
\label{planar}
In the strict limit of $k,N,M\rightarrow\infty$ we can approximate the summation by the integration over a segment:
\begin{align}
F^{\text{(mat)}}(k,N,N+M)&=2N^2\int_0^1 ds(1-s)\log 2\cos \pi\Bigl(\lambda_1s+\frac{\lambda_2-\lambda_1}{2}\Bigr)\nonumber \\
&\quad +NM\int_0^1ds\log 2\cos\pi\Bigl((\lambda_2-\lambda_1)s-\frac{\lambda_2-\lambda_1}{2}\Bigr)+\cdots.
\label{planarapprox}
\end{align}
Here the integrations can be performed explicitly by expanding $\log 2\cos z$ formally as
\begin{align}
\log 2\cos z=\sum_{n=1}^\infty\frac{(-1)^{n-1}}{2n}(e^{2inz}+e^{-2inz}),
\end{align}
which are concisely expressed with the trilogarithm.  
By using the notation of \eqref{tHooftExpansion} we obtain 
\begin{align}
&F^{\text{(mat)}}_0(\lambda_1,\lambda_2) \nonumber \\
=&\frac{1}{4\pi^2}(\Li_3(-e^{\pi i(\lambda_2+\lambda_1)})+\Li_3(-e^{-\pi i(\lambda_2+\lambda_1)})-\Li_3(-e^{\pi i(\lambda_2-\lambda_1)})-\Li_3(-e^{-\pi i(\lambda_2-\lambda_1)})).
\label{Fmatplanar}
\end{align}
Note that the leading free energy is an odd function in terms of $\lambda_1$. 

In general, the replacement $\sum_{\alpha=1}^K f(\alpha/K) \overset{K\to\infty}\rightarrow K\int_0^1 ds f(s)$ contains possible ${\cal O}(K^{-1})$ corrections in the integrand.
Hence the evaluation above is valid only for the leading part in the large $k,N,M$ limit.

\subsection{All order 't Hooft expansion}
Let us compute the $1/k$ corrections by reorganizing the summations in \eqref{rearranged} into an appropriate infinite series.

First we carry out the Taylor expansion for each $\log \cos$ as
\begin{align}
\log\cos(\pi z)=\sum_{m=1}^\infty \frac{-(2^{2m}-1)\zeta(2m)}{m}z^{2m}.
\end{align}
Changing the order of summation, we obtain
\begin{align}
&F^{\text{(mat)}}(k,N,N+M) \nn
=& N(N+M)\log 2 - \sum_{m=1}^\infty
\frac{(2^{2m}-1)\zeta(2m)}{m k^{2m} }
\Biggl[
2\sum_{\alpha=1}^N(N-\alpha)\Bigl(\alpha+\frac{M}{2}\Bigr)^{2m}
+N\sum_{\alpha=0}^M\Bigl(\alpha-\frac{M}{2}\Bigr)^{2m}
\Biggr] \nn
=& N(N+M)\log 2  - \sum_{m=1}^\infty
\frac{(2^{2m}-1)\zeta(2m)}{m k^{2m} }
\nonumber \\
&\quad \times 2\bigg[ 
\Bigl(N+{M\over2}\Bigr) I_{2m}\Bigl(N+{M\over2}\Bigr) -{M\over2} I_{2m}\Bigl({M\over2}\Bigr) - I_{2m+1}\Bigl(N+{M\over2}\Bigr) +I_{2m+1}\Bigl({M\over2}\Bigr)  \bigg],
\end{align}
where we set $I_\nu(N)=\sum_{j=1}^N j^\nu$. $I_\nu(N)$ admits the $1/N$ expansion by the Bernoulli formula 
\begin{align}
I_\nu(N)
=\sum_{\ell=0}^\nu(-1)^\ell 
{1\over \nu+1} \begin{pmatrix}
 \nu+1 \\
 \ell\\ 
\end{pmatrix} B_\ell N^{\nu+1-\ell},
\end{align}
with $B_\ell$ the Bernoulli numbers defined by $x/(e^x -1) = \sum_{\ell=0}^\infty B_\ell/\ell! x^\ell$. 
Thus we can replace the original finite summation with an infinite series, where each term in summation has definite powers of $k,N,M$.
Since $N=k\lambda_1,M=k(\lambda_2-\lambda_1)$, this directly provides the $1/k$ expansion.
Collecting the coefficients together, we finally obtain 
\begin{align}
F^{\text{(mat)}}_0(\lambda_1,\lambda_2)&=\lambda_1\lambda_2\log 2+\sum_{m=1}^\infty\frac{-(2^{2m}-1)\zeta(2m)}{m(m+1)(2m+1)}
\Bigl[
\Bigl(\frac{\lambda_2+\lambda_1}{2}\Bigr)^{2m+2}-
\Bigl(\frac{\lambda_2-\lambda_1}{2}\Bigr)^{2m+2}
\Bigr],\nonumber \\
F^{\text{(mat)}}_g(\lambda_1,\lambda_2)&=\sum_{m=0}^\infty\frac{(2^{2m+2g}-1)\zeta(2m+2g)B_{2g}}{g(m+g)}
\binom{2m+2g}{2g-2}\nonumber \\
&\quad\quad\quad\quad\times\Bigl[
\Bigl(\frac{\lambda_2+\lambda_1}{2}\Bigr)^{2m+2}-
\Bigl(\frac{\lambda_2-\lambda_1}{2}\Bigr)^{2m+2}
\Bigr],
\label{Fmatatgenusg}
\end{align}
where $g$ runs all positive integers. 
Note that although in the beginning there exist contributions with $g$ half integer, they totally vanish due to the fact that $B_{2n+1}=0$ for $n\geq 1$.
The planar part $F_0^{\text{(mat)}}(\lambda_1,\lambda_2)$ is consistent with the result \eqref{Fmatplanar} which we have obtained by replacing the summations with one dimensional integrations.

\subsection{Comments on higher Spin/Supergravity crossover}
Now we comment on the vector model limit, where $k,M\rightarrow\infty$ with $N$ and $\lambda_2$ kept finite.
This limit can be achieved by the strict 't Hooft limit with $\lambda_1\to0$.

The large $M$ scaling behavior of the free energy in the vector model limit can be deduced from the planar result in the following way.
Suppose the planar part of the free energy is expanded in terms of $\lambda_1$ around $\lambda_1 \sim 0$ as 
\begin{align}
F=k^2(\lambda_1^\nu f(\lambda_2)+{\cal O}(\lambda_1^{\nu+1}))+\cO(k).
\end{align}
Then the free energy scales in the vector model limit as 
\begin{align}
F\sim k^{2-\nu} \sim M^{2-\nu}. 
\end{align}
This is because by using $\lambda_1= N/k$, this can be written as
\begin{align}
F=k^{2-\nu}( N^\nu f(\lambda_2)+k^\nu{\cal O}(\lambda_1^{\nu+1})) +\cO(k^{1-\nu}) =k^{2-\nu} N^\nu f(\lambda_2)+\cO(k^{1-\nu}).
\end{align}

In our case, from \eqref{Fmatplanar}, the small $\lambda_1$ expansion of $F^{\text{(mat)}}(k,N,N+M)$ starts with
\begin{align}
F^{\text{(mat)}}(k,N,N+M)=k^2\(\lambda_1 \frac{i(\Li_2(-e^{\pi i\lambda_2})-\Li_2(-e^{-\pi i\lambda_2})}{2\pi}+{\cal O}(\lambda_1^3)\) +\cO(k).
\end{align}
Hence $F^{\text{(mat)}}(k,N,N+M)$ is linear in $M$ in the vector model limit, which is the expected result. 

It is expected that the $\cN=4$ Chern-Simons matter theories exhibit higher spin symmetry in the vector model limit \cite{Chang:2012kt,Jain:2012qi} (see also \cite{Giombi:2011kc,Aharony:2011jz}), 
where higher spin currents transform as the adjoint representation of U$(N)$. 
In fact, there is a conjecture that the minimal $\cN=4$ theory investigated in this paper 
corresponds to a parity-violated supersymmetric higher spin Vasiliev theory on AdS$_4$ \cite{Chang:2012kt}. Here the higher spin fields have the U$(N)$ color index, where the gauge interaction is governed by the bulk 't Hooft coupling $(N+M)/N$.
It was argued that at the weak 't Hooft coupling region ($N\sim\infty$) the theory is well described by a certain colored supergravity, while at the strong bulk 't Hooft coupling ($N\sim1$) the bulk color interaction is confined so that the theory is described by the higher spin Vasiliev theory. 
In the case of ABJ theory \cite{Aharony:2008gk} this transition was shown to occur in a smooth fashion \cite{Hirano:2015yha}. 
We have showed that this is also the case in the minimal Gaiotto-Witten theory.

\section{Discussion}
\label{Discuss}

In this paper we have computed the partition function of the ${\cal N}=4$ $\text{U}(N)_k\times \text{U}(N+M)_{-k}$ linear quiver superconformal Chern-Simons theory on three sphere.
By performing the $2N+M$ integrations in the localization formula \eqref{localization} explicitly, we have obtained the closed form expression for the partition function \eqref{factorize}.
Interestingly, we have found that the factor $Z^{\text{(mat)}}_{\zeta_1+\zeta_2}(k,N,N+M)$ in \eqref{factorize}, which is a priori mere the remainder after the two pure Chern-Simons partition function, takes completely the same form as the one-loop determinant of the bifundamental hypermultiplet in the localization formula \eqref{localization}.
This motivates us to refer to our result as ``complete factorization''.
The closed form expression has also enabled us to show the invariance of the partition function under the level/rank duality, which was also confirmed from the Hanany-Witten transition in the type IIB brane configuration, and to compute the all order 't Hooft explanation of the free energy.

To understand the implication of ``factorization'', it is important to assign more concrete interpretation to $Z^{\text{(mat)}}$.
One tempting candidate may be to regard the whole partition function as the pure Chern-Simons partition function for the supergroup $\text{U}(N|N+M)$.
In fact, the partition function of pure Chern-Simons theory appears as a modular transformation matrix of the characters in the corresponding Affine Lie algebra associated with the gauge group \cite{Witten:1988hf}.
Our simple result of the matter partition function \eqref{factorize} may have such a purely algebraic and representational origin. It would be interesting to elucidate the origin by generalizing modular transformation matrices of general Affine Lie algebras classified in \cite{Kac:1984mq} including matter contribution.

Although our result \eqref{factorize}-\eqref{Zmat} may be valid only in the region $N+M/2-1<|k|/2$, the result is meaningful beyond this region after the analytic continuation in $k$, which is indeed necessary to argue the level-rank duality.
Once we define the partition function by the analytic continuation, we find that the partition function has poles at \eqref{polecondition}.
This singular behavior is opposite from what happens in the violation of the bounds for the s-rule: the vanishing of the partition function which one might associate with the supersymmetry breaking in analogy of the argument with the Witten index on $S^2\times S^1$.
It will be interesting to provide a physical interpretation for these poles.\footnote{
We have also observed that we can trace the divergence starting from a mass deformed ABJM theory which reduces to the Gaiotto-Witten theory in the decoupling limit.
This result will be reported in \cite{NYfuture_bound}.
}

An analogous divergence is indeed known for the ${\cal N}=4$ $\text{U}(N)$ gauge theory (without Chern-Simons term) coupled with $N_f$ matter multiplets, where the $S^3$ partition function diverges for $N_f/2<N_c\le N_f$ \cite{Kapustin:2010mh,Yaakov:2013fza}.
In this case the theory is ``bad'' for $N_f/2<N_c\le N_f$ and can be dual to a ``good'' theory with decoupled massless hypermultiplets.
The contribution from the decoupled sector appears as a prefactor in the duality relation, which explains the divergence in the bad theory.
The divergence in our theory as well as the relation under the level/rank exchange \eqref{GWlevelrank} might be interpreted in the same manner.
To clarify this idea it would be important to identify the unitarity violating monopole operators or show their existence from e.g. the computation of the superconformal index of the $\text{U}(N)_k\times \text{U}(N+M)_{-k}$ linear quiver theory.

It is also interesting to compare our results with the scattering amplitudes in quantum field theories, where the analytic continuation is necessary as well to see their crossing symmetry.
Indeed, a similar kind of prefactor as \eqref{GWlevelrank} appears in the crossing symmetry relation of scattering amplitude in Chern-Simons vector theory  \cite{Jain:2014nza,Inbasekar:2015tsa,Yokoyama:2016sbx}.
That prefactor is important not only for crossing symmetry but also for unitarity in scattering amplitudes.    
After noticing the similarity, it would be interesting to interpret the poles \eqref{polecondition} in analogy to the bound states in scattering amplitude.

In \cite{Gopakumar:1998jq} the free energy of the pure Chern-Simons theory was found to coincide with the topological string on the resolved conifold in the 't Hooft expansion.
It would be fascinating to provide a topological string interpretation to \eqref{Fmatatgenusg} as well. 
The obstruction of this program will be in the fermionic nature of the matter partition function \eqref{factorize} in contrast to the bosonic nature of that of pure Chern-Simons theory. 
This difference is usually crucial in the free energy, which can be seen for example in cancellation of the contribution of high energy modes. 
Reproducing the matter partition function from the topological string may require some new topological object.

The minimal Gaiotto-Witten theory we have considered can be obtained from the ABJM theory by removing one bifundamental hypermultiplet and cutting open the circular quiver.
Indeed, our computation in \S\ref{ExactPf} for $M=0$ can be reorganized as the ``open version'' of the Fermi gas formalism \cite{Marino:2011eh}, where we compute the matrix elements over only $N$ distinctive one-particle states instead of the trace over whole Hilbert space of $N$ fermions.
Once we notice this viewpoint we can perform similar computation also for the linear quivers arising from the brane setup \eqref{GWN4IIB} with additional $(1,k')$5-branes as well, where $k'$ is not necessarily equal to $k$.
We shall report the detailed analysis for these generalizations, together with the rank deformations and the addition of fundamental matter fields, in a separated work \cite{NYfuture_topologicalstring}.

We hope to report any progress on these issues in near future. 

\appendix

\section{Determinant Formulas}
\label{A_Formulas}
\begin{itemize}

\item Cauchy-Vandermonde determinant formula:
\begin{align}
\frac{
\displaystyle
\prod_{a<b}^N
(x_a-x_b)
\prod_{i<j}^{N+M}
(y_i-y_j)
}
{
\displaystyle \prod_{a=1}^N\prod_{j=1}^{N+M}(x_a+y_j)
}
=(-1)^{MN+\frac{M(M-1)}{2}}\determinant{(a\oplus \ell),j}
\begin{pmatrix}
\displaystyle \biggl[\frac{1}{x_a+y_j}\biggr]_{N\times (N+M)}\\
\displaystyle \Bigl[y_j^{\ell-1}\Bigr]_{M\times (N+M)}
\end{pmatrix}.
\label{CauchyVdM}
\end{align}
\item Convolution of determinants (formula (A.1) in \cite{Matsumoto:2013nya})
\begin{align}
\frac{1}{N!}\int dx^N\determinant{a,b}\biggl[\psi_a(x_b)\biggr]\determinant{(a\oplus \ell),j}
\begin{pmatrix}
\displaystyle \Bigl[\phi_j(x_a)\Bigr]_{N\times (N+M)}\\
\displaystyle \Bigl[\xi_{\ell,j}\Bigr]_{M\times (N+M)}
\end{pmatrix}
=
\determinant{(a\oplus \ell),j}
\begin{pmatrix}
\displaystyle\Bigl[\int dx\psi_a(x)\phi_j(x)\Bigr]_{N\times (N+M)}\\
\displaystyle\Bigl[\xi_{\ell,j}\Bigr]_{M\times (N+M)}
\end{pmatrix}.
\label{detdet}
\end{align}
\end{itemize}

\section{Comment on computation of ${\mathscr M}_{1,a,j}$}
\label{commentonM1}

In this section we revisit the computation of the matrix element ${\mathscr M}_{1,a,j}$ \eqref{calM1result}.
In the computation in \S\ref{CompleteFactor} we have formally applied the following integration formula to the $\mu_+$-integration to produce the delta function
\begin{align}
\int_{-\infty}^{\infty}ds e^{its}=2\pi\delta(t),\quad (t\in\mathbb{R}).
\label{fourierdelta}
\end{align}
Though the choice of the integration contour $s\in (-\infty,\infty)$ together with the reality of $t$ is crucial in this formula, these assumptions are not satisfied in the original integration \eqref{matrixelements}.

Below we explain the computation in more detail.
After a trivial change of the integration variables $(\mu,\nu)\rightarrow (x,y)=(\mu+\nu,\mu-\nu)$, \eqref{matrixelements} is written as
\begin{align}
{\mathscr M}_{1,a,j}=\frac{1}{8\pi^2}\int_{-\infty}^{\infty} dx\int_{-\infty}^\infty dy\frac{1}{2\cosh\frac{y}{2}}e^{\frac{ikxy}{4\pi}+Ax+By},
\label{M1inAB}
\end{align}
where
\begin{align}
A=\frac{1}{2}\Bigl[a-j+\frac{M}{2}+i(\zeta_1+\zeta_2)\Bigr],\quad
B=\frac{1}{2}\Bigl[a+j-N-1+\frac{M}{2}+i(\zeta_1-\zeta_2)\Bigr].
\label{AB}
\end{align}
Starting from \eqref{M1inAB} we shall follow the following two different ways: deform of integration contour so that the formula \eqref{fourierdelta} is applicable (\S\ref{contour}) and change the order of integration so that the computation does not involve delta functions (\S\ref{coshcoshtwice}).
Although giving two explanations would be redundant, we hope they are helpful in future for a more rigorous justification or generalizations of the model.
Also note that both of the two computations assume $|\text{Re}[B]|<1/2$ as a regularization, whose justification is a subtle issue.\footnote{
In order the assumption to make sense, here we should regard $B$ in ${\mathscr M}_{1,a,j}$ with different $(a,j)$ as independent variables.
Otherwise, since $(a,j)$ takes values of $a=1,2,\cdots,N$ and $j=1,2,\cdots,N+M$, the assumption $|\text{Re}[B]|<1/2$ requires $N+3M/2<2$, which is more strict than \eqref{noresiduebound}.
% The condition $N+3M/2<2$ indeed corresponds to the condition for the absolute convergence of the matrix model \eqref{localization}.
We observe, however, that our results for the partition function \eqref{factorize}-\eqref{Zmat} are indeed obtained even for $N+3M/2\ge 2$ if we define $Z_{\zeta_1,\zeta_2}(k,N,N+M)$ as the decoupling limit of the mass deformed ABJM matrix model \cite{NYfuture_bound}.
We hope to provide a more rigorous formulation for the partition function in \cite{NYfuture_bound}.
}

\subsection{Deformation of integration contour}
\label{contour}

In \S\ref{CompleteFactor} we have naively applied the integration formula \eqref{fourierdelta} to the $x$-integration in the expression \eqref{M1inAB}.
Here let us look this step carefully.
In order to apply the formula \eqref{fourierdelta}, we have to deform the contour of $y$-integration so that $ky/4\pi-iA\in\mathbb{R}$.
For this purpose we denote $y$ as $y=z+4\pi iA/k$ and decompose the $z$-integration as
\begin{align}
{\mathscr M}_{1,a,j}&=\frac{1}{8\pi^2}\int_{-\infty}^{\infty} dx\int_{-\infty-\frac{4\pi i\text{Re}[A]}{k}}^{\infty-\frac{4\pi i\text{Re}[A]}{k}}dz\frac{e^{Bz}}{2\cosh\frac{1}{2}(z+\frac{4\pi iA}{k})}e^{\frac{ikxz}{4\pi}}\nonumber \\
&=\frac{e^{\frac{4\pi iAB}{k}}}{8\pi^2}\int_{-\infty}^{\infty} dx\biggl[
\int_{-\infty}^{\infty}dz\frac{e^{i(\frac{kx}{4\pi}-iB)z}}{2\cosh\frac{1}{2}(z+\frac{4\pi iA}{k})}
+\int_{\gamma}dz\frac{e^{i(\frac{kx}{4\pi}-iB)z}}{2\cosh\frac{1}{2}(z+\frac{4\pi iA}{k})}
\biggr],
\label{deformdecompose}
\end{align}
where the integration contour $\gamma$ is given as follows
\begin{align}
\includegraphics[scale=0.3]{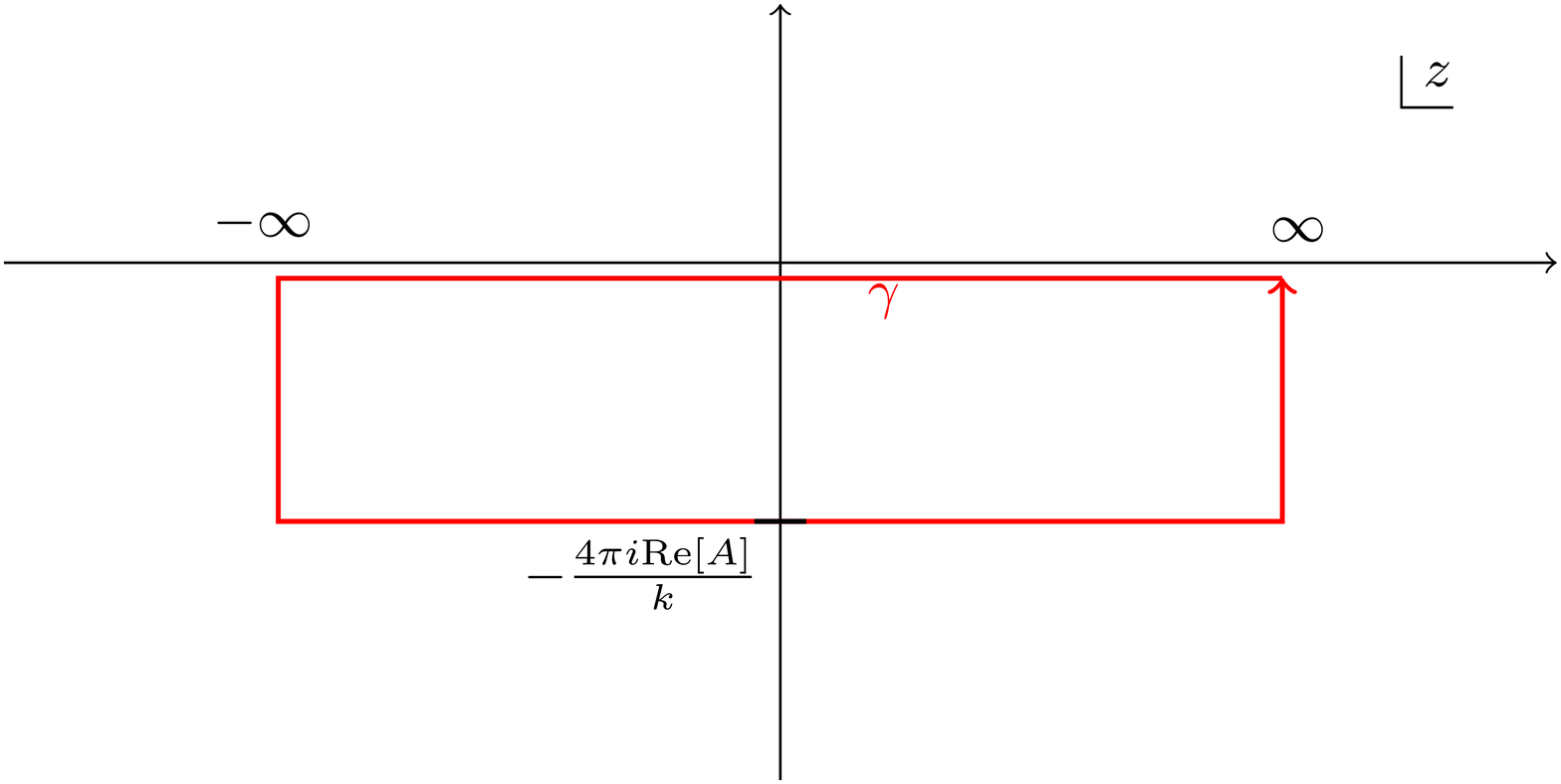}.
\label{gamma123}
\end{align}
If the contributions from $\gamma$ vanishes, we can perform the $x$-integration in the first term and obtain
\begin{align}
{\mathscr M}_{1,a,j}&=\frac{e^{\frac{4\pi iAB}{k}}}{4\pi}\int_{-\infty}^{\infty}dz\frac{e^{Bz}}{2\cosh\frac{1}{2}(z+\frac{4\pi iA}{k})}\delta\Bigl(\frac{kz}{4\pi}\Bigr)\nonumber \\
&=\frac{e^{\frac{4\pi iAB}{k}}}{2k\cosh\frac{2\pi iA}{k}},
\label{M1inAB_result}
\end{align}
which reproduces the result in \S\ref{CompleteFactor} \eqref{calM1result}.

Hence our computation in \S\ref{CompleteFactor} is rigorous so far only when the $z$-integration in \eqref{deformdecompose} over $\gamma$ vanishes.
Since the integrand in \eqref{deformdecompose} has poles at $z=\pi in-4\pi iA/k$ with $n\in\mathbb{Z}$, this requires $|\text{Re}[A]|<|k|/4$.
As the real part of $A$ \eqref{AB} is maximized as $\text{Re}[A]=(N+M/2-1)/2$ at $(a,j)=(1,N+M)$ and minimized as $\text{Re}[A]=-(N+M/2-1)/2$ at $(a,j)=(N,1)$, this condition is guaranteed when
\begin{align}
N+\frac{M}{2}-1<\frac{|k|}{2}.
\label{noresiduebound}
\end{align}

\subsection{Computation without delta function}
\label{coshcoshtwice}
We can also compute ${\mathscr M}_{1,a,j}$, without explicit appearance of a delta function, with the help of the following Fourier transformation formula
\begin{align}
\int_{-\infty}^{\infty}ds\frac{e^{its}}{2\cosh\frac{s}{2}}=\frac{\pi}{\cosh\pi t},
\label{fourierselfdual}
\end{align}
which is convergent if $|\text{Im}[t]|<1/2$.
Applying this formula to the $y$-integration in \eqref{M1inAB} (the constraint $|\text{Im}[t]|<1/2$, with $t=kx/(4\pi)-iB$, is satisfied due to the assumption $|\text{Re}[B]|<1/2$) we obtain
\begin{align}
{\mathscr M}_{1,a,j}=
\frac{1}{8\pi}
\int_{-\infty}^{\infty}dx
\frac{e^{Ax}}{\cosh(\frac{kx}{4}-\pi iB)}
=
\frac{e^{\frac{4\pi iAB}{k}}}{8\pi}
\int_{-\infty+\pi i \text{Re}[B]}^{\infty+\pi i \text{Re}[B]}dz
\frac{e^{\frac{2Az}{k}}}{\cosh\frac{z}{2}}
=
\frac{e^{\frac{4\pi iAB}{k}}}{8\pi}
\int_{-\infty}^{\infty}dz
\frac{e^{\frac{2Az}{k}}}{\cosh\frac{z}{2}},
\end{align}
where $z=kx/2+\pi iB$.
Here we have deformed the integral contour so as to pass through the origin without picking up any pole due to the assumption $|\text{Re}[B]|<1/2$.
Now we encounter the same constraint $|\text{Re}[A]|<|k|/4$, or \eqref{noresiduebound}, which is required here for the convergence of $z$-integration.
When the condition \eqref{noresiduebound} is satisfied we can apply the formula again \eqref{fourierselfdual} to obtain the same expression for ${\mathscr M}_{1,a,j}$ as \eqref{M1inAB_result}.

\section*{Acknowledgement}
We would like to thank Masazumi Honda, Sanefumi Moriyama, Takao Suyama, Minwoo Suh, Tadashi Okazaki, Sungjay Lee and Itamar Yaakov for valuable comments.

%%%%%%%%%%%%%%%%%%%%%%%%%%%%%%%%%%%%%%%%%%%%%%%%%%%%%%%%
\bibliographystyle{utphys}
\bibliography{GWFactorizev4}
\end{document}